%Paper: hep-ph/9304299
%From: leigh@scipp.UCSC.EDU
%Date: 26 Apr 1993 18:46:52 -0700 (PDT)
%Date (revised): 07 May 1993 23:04:31 -0700 (PDT)

\input phyzzx
\input tables
%\PHYSREV
%\input twocol
%
% Some useful macros
\def\SCIPP{\centerline {\it Santa Cruz Institute for Particle Physics}
  \centerline{\it University of California, Santa Cruz, CA 95064}}
\def\SLACT{\centerline {\it Stanford Linear Accelerator Center}
  \centerline{\it Stanford University, Stanford, CA 94309}}
\def\eg{{\it e.g.}}
\def\ie{{\it i.e.}}
\def\etc{{\it etc.}}
\overfullrule 0pt
\vskip1cm
\Pubnum{SCIPP 93/04 \cr
SLAC-PUB-6147}
\date{March, 1993}
\titlepage
\pubtype{ T}     % T, E, or T/E  (Theory or Experimental)
\vskip2cm
\title{Flavor Symmetries and The Problem of Squark Degeneracy}
\author{Michael Dine and Robert Leigh}
\address{\SCIPP}
\vskip.2cm
\author{Alex Kagan}
\address{\SLACT}
\vskip1.5cm
\vbox{
\centerline{\bf Abstract}
If supersymmetry exists at low energies, it is necessary to understand
why the squark spectrum exhibits sufficient degeneracy to suppress
flavor changing neutral currents.  In this note, we point out that
gauged horizontal symmetries can yield realistic quark mass matrices,
while at the same time giving just barely enough squark degeneracy to
account for neutral $K$-meson phenomenology.  This approach
suggests likely patterns for squark masses, and indicates that
there could be significant supersymmetric contributions to
$B-\bar{B}$ and $D-\bar{D}$ mixing and $CP$-violation in the $K$ and $B$
 systems.
}
\submit{Physical Review D}

\parskip 0pt
\parindent 25pt
\overfullrule=0pt
\baselineskip=18pt
\tolerance 3500

\endpage
%\pagenumber=1
%\chapter{Introduction}

\REF\randall{
For recent efforts to solve these problems, see
L. Randall, MIT preprint MIT-CTP-2112 (1992) and H. Georgi,
Harvard preprint HUTP-92-1037 (1992).}
\REF\nilles{H.P. Nilles, {\sl Phys. Rep.} {\bf 110} (1984) 1.}
\REF\masiero{F. Gabiani and A. Masiero, {\sl Nucl. Phys.} {\bf B322}
(1989) 235}
\REF\dinenelson{For a recent attempt
along these lines, see M. Dine and A. Nelson, UCSD preprint UCSD/PTH
93-05 (1993).}
\REF\critique{H. Georgi, {\sl Phys. Lett.} {\bf 169B} (1986) 231;
L.J. Hall, V.A. Kostelecky and S. Raby, {\sl Nucl. Phys.} {\bf B267}
(1986) 415.}
\REF\hagelin{J.S. Hagelin, S. Kelley, and T. Tanaka,
MIU preprint MIU-THP-92/59.}
\REF\dks{M. Dine, A. Kagan and S. Samuel, {\sl Phys. Lett.} {\bf 243B} (1990)
250.}
\REF\ibanez{L. Ibanez and D. Lust, {\sl Nucl. Phys.} {\bf B382}
(1992) 305; B. de Carlos, J.A. Casas and C. Munoz, CERN-TH.6436/92 and
{\sl Phys. Lett.} {\bf 299B} (1993) 234}
\REF\vadim{V.S. Kaplunovsky and J. Louis, preprint
CERN-TH. 6809/93 UTTG-05-93.}

Two solutions of the hierarchy problem have been suggested over the
years:  technicolor and supersymmetry.  Perhaps the biggest problem
for technicolor theories is that they tend to suffer from unacceptable
flavor-changing neutral currents.  Partial solutions to this problem
have been offered, but the resulting models are
extremely elaborate.\refmark{\randall}  Supersymmetry, it is often argued,
does not suffer from this problem.  However, this is not so clear.
At one loop, it is well known that there are diagrams contributing
to $K^o -\bar K^o$ mixing which, for supersymmetry breaking masses
below a TeV, are too large unless there
is a high degree of degeneracy among squarks.
The real part of this mixing, for example, leads to the requirement
that\refmark{\nilles,\masiero,\hagelin}
$${\delta \tilde m_q^2 \over m_{susy}^2}
{\delta \tilde m_{\bar {q}}^2 \over m_{susy}^2}{1 \over
m_{susy}^2}\lsim ~10^{-10}~ {\rm GeV}^{-2}
\eqn\degenlimit$$
while for the imaginary part, the limit is about two orders
of magnitude stronger.  There are also limits on degeneracy from other
processes:  $B -\bar B$ mixing, $b\rightarrow s \gamma$,
$\mu \to e \gamma$, \etc\ There are additional
constraints on the size of certain $CP$-violating angles coming from
$K-\bar K$, and the neutron and electron electric dipole moments.
The question is, can one naturally satisfy all of these constraints?   In some
early models of supersymmetry breaking, these conditions were automatically
satisfied because the breaking of supersymmetry was fed to squarks
by gauge interactions.\refmark{\dinenelson}
In hidden sector supergravity theories,
however, which provide the basis for much of our thinking about low energy
supersymmetry, the situation is far less clear.  It is often said that
this degeneracy is perhaps reasonable, since, after all, gravity is
``flavor blind."  On closer examination, however, this argument is seen
to be without substance.  In most
models of the type which have been
considered to date, there are operators which one can add to
the theory, not suppressed by any (even approximate) symmetry,
which give rise to an ${\cal O}(1)$ breaking of the degeneracy.
This problem has been discussed in numerous places.  In the
context of supergravity theories, for example, it is considered
in ref. \critique. In ref. \dks,  this situation was anticipated for
string theory and
strategies for naturally raising the supersymmetry breaking scale into the
multi-TeV region to alleviate
this problem were proposed.   Explicit departures from universality
in simple orbifold models have been computed
in ref. \ibanez.  Kaplunovsky and Louis \refmark{\vadim} recently
have reviewed
this problem in the framework of string theory.  They note that if
supersymmetry breaking is associated principally with the dilaton,
one will obtain some degree of degeneracy.  However, they have
also pointed out serious difficulties with such a scenario.

\REF\whl{L. Hall, J. Lykken and S. Weinberg, {\sl Phys. Rev.} {\bf D27}
(1983) 2359.}
\REF\banksdixon{T. Banks and L. Dixon, {\sl Nucl. Phys.} {\bf B307}
(1988) 93.}
In early work on hidden sector supergravity models,
it was suggested that
one should simply postulate a large, approximate,
flavor symmetry among squarks.\refmark{\whl}
Indeed, while various other solutions to this problem might
be contemplated, flavor symmetries seem a most natural framework.
There are two immediate issues which one must face.  First, whatever
horizontal symmetry there may be is clearly very badly broken
by the ordinary quark mass matrices.  Second, we would prefer not
to impose continuous global symmetries on the underlying theory.  Such
symmetries are almost certain to be broken by gravitational interactions,
and are known not to arise in string theories.\refmark{\banksdixon}

In this paper we will study models with non-abelian, gauged
flavor symmetries, to determine whether these can assure
an adequate degree of squark degeneracy while simultaneously
allowing realistic quark mass matrices.
We will describe simple models containing an $SU(2)_H$ horizontal
symmetry in which there is adequate degeneracy to satisfy the
limits coming from the real part of $K$-$\bar K$
mixing.  To be more precise, a naive estimate, assuming all
susy breaking parameters of order 300 GeV, gives a result about
an order of magnitude larger than the experimental upper bound.
This order of magnitude discrepancy is not disturbing.  First,
in the framework we consider, it is not unnatural to
suppose that squarks of the first generation have TeV masses,
while those of the third have smaller masses (so fine tuning of Higgs
masses is not required).  Alternatively, some of the parameters
of order one in the model may be of order $1/3-1/10$.
The limit on the
imaginary part, two orders of magnitude stronger, is more problematic.
%The simplest models and assumptions give a degree of degeneracy two to
%three orders of magnitude too large.
To satisfy this constraint, it is necessary to make some
further assumptions.  Again, there are plausible regions of parameter
space for which the imaginary part is sufficiently small.
%  First, one can consider certain
%special regions of parameter space.  In the models we
%will consider, it is not unnatural for the squarks
%of the first two generations to have $TeV$ type masses,
%while the third generation squarks have significantly
%smaller masses; this can ameliorate these problems.  Alternatively,
%one can take
%certain dimensionless parameters, nominally of order one,
%to be $\sim .1$.
A different approach is to impose
additional symmetries, such as discrete
symmetries, to provide further suppression.
This seems a reasonable thing to do, since
such symmetries might be necessary to understand
\REF\nir{M. Leurer, Y. Nir and N. Seiberg, Rutgers preprint RU-92-59, (1992).}
the fermion mass matrices.\foot{For a recent effort along
these lines, see ref. \nir.}
One also may want to consider
additional assumptions about the nature of $CP$-violation.  Note that
bounds on gaugino mass phases from $d_n$ and $d_e$
also suggest additional assumptions such as spontaneous $CP$-violation.
One can  view these results in a positive light:  generic
models do not (quite) satisfy all constraints, so additional features
must be considered -- and perhaps additional predictions made.

The models we consider will be predictive:  they will
imply definite relations among squark masses.
For example, models with $SU(2)_H$ symmetry predict
that up or down squarks of the first two families are
approximately degenerate,
while third family squarks may have quite different masses.
Similar degeneracies among sleptons are also expected.
These models will also have interesting implications for
$B-\bar B$ and $D-\bar D$ mixing and, possibly, $b \to s \gamma$.

One might hope that models of this
kind would explain the many puzzling
features of the fermion mass spectrum.
We will not attempt this here.  In particular, our models will
require a rather large range of quark Yukawa couplings (though
perhaps
not quite as large a range as in the minimal standard model).

The first question one must address is the scale of breaking
of the horizontal symmetry.  We will distinguish
two possibilities:  breaking near $M_p$,
and breaking much below $M_p$.
A simple model with large-scale breaking is the following.  Take the
gauge group to be that of the standard model times an additional
$SU(2)_H$.  For purposes of enumerating the different particles
and couplings, we will label the states by the quantum numbers
they might have in an $SU(5) \times SU(2)_H$ unification.
Note that we are not assuming an underlying $SU(5)$ symmetry,
but simply using $SU(5)$ to classify the states.
The three generations are then assumed to form doublets
and singlets of the $SU(2)_H$.  The states are
$$\bar 5_a= (\bar 5, 2) ~~~~10_a = (10,2)~~~~
\bar 5_s = (\bar 5, 1) ~~~~10_s = (10,1). $$
The Higgs particles are taken to be two  singlets of $SU(2)_H$;
this will avoid the problem of flavor changing currents mediated
by Higgs particles.
We will denote these by $H_1$ and $H_2$.
To break $SU(2)_H$, one adds
three fields transforming as $(1,2)$:
$\Phi_i^a,i=1,2,3$.  The model
is then free of both perturbative and
non-perturbative anomalies.  Alternatively, one can add an $SU(2)_H$
doublet of right-handed neutrinos, in which case only two $\Phi _i$
singlets are added.

\REF\worldsheet{M. Dine, N. Seiberg, X.G. Wen and E. Witten,
{\sl Nucl. Phys.} {\bf B278} (1986) 769;
L. Dixon, in {\it Superstrings, Unified
Theories, and Cosmology 1987}, G. Furlan, et al, eds., World Scientific
(Singapore, 1988).}
\REF\discreter{M. Dine and N. Seiberg, {\sl Nucl. Phys.} {\bf B306}
(1988) 137.}
\REF\fayet{M. Dine, N. Seiberg and E. Witten, {\sl Nucl. Phys.}
{\bf B289} (1987) 589.}
We will assume that supersymmetry is broken in a hidden sector,
whose dynamics do not by themselves break any of these
gauge symmetries.    We will also assume that, after supersymmetry
breaking, the potential for the fields $\Phi_i$ is such that these
fields obtain large vev's.  This assumption may
seem unnatural, but it is often
satisfied in string theories. First, there are ``D-flat" directions
(\ie, directions in which the auxiliary $D$-fields
in the $SU(2)_H$ gauge supermultiplet vanish)
where some of these fields have vev's. $F$-flatness is known to
arise in string theory in at least two ways.
Moduli of string compactifications with $(2,2)$ world
sheet supersymmetry are $F$-flat.\refmark{\worldsheet}
At points of enhanced
symmetry, the moduli are typically charged (\eg, at orbifold
points); the enhanced symmetry could be our horizontal
symmetry.  Generically, however, the moduli {\it do} appear in the
superpotential of the matter fields (some of these couplings
may be exponentially suppressed at large radius).\refmark{\worldsheet}
$F$-flatness is also
known to arise in the presence of discrete
$R$-symmetries.\refmark{\discreter}
In either case,
if some of the $\Phi$ fields acquire negative masses upon
supersymmetry breaking, they can acquire large vev's.  We will
require that these be smaller than $M_p$ by a factor of order $10$.
We will not attempt here to explain how this factor might arise, but
simply argue that in a theory with small couplings it is not unnatural.
\foot{One possible origin of this scale is through the appearance
of a Fayet-Iliopoulos D-term.\refmark{\fayet}}

To keep the discussion simple, we will assume that two singlets,
$\Phi_1$ and $\Phi_2$, obtain vev's:
$\vert \langle \Phi_1  \rangle
\vert =(0,\phi)^T$ and $\vert \langle \Phi_2 \rangle \vert =(\phi,0)^T$.
\foot{The precise alignment of $\Phi_1$ and $\Phi_2$
will not be important to us, except when we consider additional discrete
symmetries.  In such cases, the alignment considered here is natural.}
In order to understand how this breaking of the $SU(2)_H$ symmetry
feeds down to other fields, we need to examine the lagrangian more
carefully. Let us focus first on the quark fields.
Denoting quark doublets by $Q$ and singlets by
$\bar d$ and $\bar u$, the superpotential just below $M_p$ contains
dimension-four terms:
$$W_q= \lambda_1 \epsilon _{ab} Q_a \bar d_b H_1 + \lambda_2 \epsilon _{ab} Q_a
\bar u_b H_2 + \lambda_3 Q_s \bar d_s H_1
+ \lambda_4 Q_s \bar u_s H_2 \eqn\wdimensionfour$$
These give rise to $SU(2)_H$ symmetric terms in the mass matrix.
Clearly we need to assume that $\lambda_1$ and $\lambda_2$
are small (this might be arranged by means of a discrete symmetry).

$SU(2)_H$-violating terms arise from higher dimension couplings,
of which there are a wide variety.  For example, in the $d$-quark
sector, we have:
$$
{1 \over M_p}(\lambda_5^i \epsilon _{ab}\Phi_a^i  Q_b \bar d_s H_1
+ \lambda_6^i \epsilon _{ab}\Phi_a^i  Q_s \bar d_b H_1)
+{1 \over M_p^2}(\lambda_7^{ij} \epsilon _{ab}\epsilon _{cd}\Phi_a^i \Phi_c^j
Q_b \bar d_d H_1 +.....)
\eqn\dimfiveandsix$$
Similar terms are present in the up-quark sector.  Some points should be noted
immediately.  First, $\Phi/M_p$ (times coupling constants)
cannot be too small; in the
limit that $\Phi \rightarrow 0$, there is no mixing of the third
generation with the first two.  As we remarked above, the $SU(2)$
symmetric terms in the light quark matrices must be small,
so $m_c$ and $m_s$ must go as
$\Phi^2 /M_p^2$; this quantity thus cannot be much smaller
than $10^{-2}$.  With this restriction on
$\Phi$, there is no difficulty in obtaining reasonable fermion masses and KM
angles, provided one is willing to take several Yukawa
couplings to be small and comparable, as in the standard
model.  Hopefully, other horizontal schemes could be more
predictive.  We leave
the exploration of this question to future work.

For simplicity,\foot{If,
for example, $m^d_{31} \sim m^d_{32}\sim m_b$,
then the mixings of the first and second generation $\bar d$ quarks
with the third generation quarks are large, and the $SU(2)$ symmetry
will not lead to sufficient degeneracy for the $K-\bar K$ system.}
 we choose the various Yukawa couplings
so that the down quark mass matrix entries $m^d_{ij} d_i \bar d_j $ satisfy
$$m^d=\left(\matrix{\sim m_d & \sim m_d & \sim m_d \cr
\sim m_d & \sim m_s & \sim m_s \cr \sim m_d & \sim m_s & \sim m_b
}\right), \eqn\downmatrix$$
with similar assumptions for the up-quark matrix.
The quark masses and
eigenstates then take on a particularly simple form.
For example, the down masses are given by
$$ m_d = m_{11}-{m_{12} m_{21} \over m_s},~~m_s = m_{22}
-{m_{23} m_{32} \over m_b},~~m_b=m_{33}. \eqn\downmasses$$

The down mass eigenstates are given by
$$ |d_i\rangle = x_{ij}^d |j \rangle, ~~~~|\bar d_i \rangle= \bar x^d_{ij}
|\bar j\rangle, \eqn\downeigenstates$$
where $i=1,2,3$ correspond to $d$, $s$, $b$, respectively, and $|1\rangle$
corresponds to the vector $(1,0,0)^T$, \etc\ With a mass matrix of the
form of eq. \downmatrix, the $x^d_{ij}$ are given by
$$x^d_{ii}\sim 1,~~x^d_{21}\sim {m^d_{12} \over m_s} \sim V_{us},~~
x^d_{32} \sim {m^d_{23} \over m_b} \sim V_{cb},~~x^d_{31}\sim
{m^d_{13}\over m_b} \sim V_{ub}. \eqn\xij$$
The remaining $x_{ij}$ follow from orthonormality of the eigenstates.
The $\bar x_{ij}$ are obtained by complex conjugating the
above and interchanging indices on the $m_{ij}$. Expressions for the
up masses and eigenstates are completely analogous.
Knowledge of the $x_{ij}$ and $\bar x_{ij}$ will be required to
estimate the various off-diagonal squark mass matrix entries of
relevance to FCNC's.

\REF\cleo{E. Thorndike, CLEO Collaboration, talk given
at the meeting of the APS, Washington, DC (1993).}
Note that, given eq. \downmatrix\  and it's analogue for the up sector,
the KM angles are essentially generated in the down sector.
The $(32)$ and $(31)$ entries in eq. \downmatrix\ are
unrelated to the KM angles and, in general, can be as large as a few GeV.
We will see that in this limit gluino graphs can give
${\rm Br}(b\to s\gamma)$ at
the level of the latest CLEO\refmark{\cleo} bound, $5.4 \times 10^{-4}$.
However, in this case we will have difficulty with the $K-\bar K$ constraints.

\REF\cargese{G. 't Hooft, in {\it Recent Developments
in Gauge Theories}, G. 't Hooft et. al, Eds., Plenum
(New York) 1980.}
What about the squark mass matrices?
We are assuming that the underlying supergravity theory is the most general
one consistent with its symmetries.  Such a theory is described, in general,
by three functions, the K\"{a}hler potential, K, the
superpotential $W$ (which we have already discussed), and a function $f$ which
describes the gauge couplings.
With our assumptions, the K\"{a}hler potential
is not of the so-called ``minimal'' type, and will give rise to
violations of universality.  If we denote ``hidden sector fields'' generically
by $z$ and visible sector fields by $y$, we can characterize the violations
of degeneracy and proportionality quite precisely.  For small
$y$, we can expand $K$ in powers of $y$.  Rescaling the fields,
we can write
$$K=k(z,z^*)+ y_i^* y_i + \ell_{ij} (z,z^*)\; y_i^* y_j+...\eqn\generalKahler$$
Examining the form of the potential in such a theory, it is easy to see that
proportionality and degeneracy occur if $\ell$ is proportional to the unit
matrix. There is no reason for this to occur in general.
However, the $SU(2)_H$ symmetry
significantly restricts the form of $\ell$.  Expanding $\ell$ in powers of
$\phi$, the leading terms for the squark fields lead to
$SU(2)_H$-symmetric squark mass terms
%  For general (non-minimal) supergravity,
%\ie, for the most general K\"{a}hler potential invariant under the
%symmetries of the model, most soft supersymmetry breaking terms allowed by
%$SU(2)_H$ will arise.  Among these will be terms
of the form (using the same
symbol for the scalar field as for the superfield)
$$V_{soft}= \tilde m_1^2 \vert Q_a \vert^2 + \tilde m_2^2 \vert Q_s \vert^2+
\tilde m_3^2 \vert \bar u_a \vert^2 +
\tilde m_4^2 \vert \bar u_s \vert^2 + ...$$
$$+A_1 \lambda_1 Q \bar d  H_1+A_2 \lambda_2 Q \bar u H_2 + .... + {\rm h.c.},
\eqn\softterms$$
Here, $\tilde m_i$ and $A_i$ are of order $m_{susy}$.
Terms linear and quadratic in $\Phi$
give rise to symmetry-breaking terms of the type:
$$\delta V^2_{soft}= {m_{susy}^2 \over M_p}(\gamma _1
\Phi _1 Q Q_s^* + ...)
+ {m_{susy}^2 \over M_p^2}(\gamma^{\prime}_1
\Phi_1 Q \Phi_2 Q^* + ...) \eqn\nonrensoftbil$$
and
$$\delta V^3_{soft} =  {m_{susy}\over M_p}
\lambda_5^1 Q \bar d_s  H_1 (\eta_1 \Phi_1
+  \eta _2 \Phi_2+ \eta_3 \Phi_2^*)$$
$$ +{m_{susy}\over M_p^2} \lambda_7^{11} Q \bar d H_1
(\eta '_1 \Phi_1 \Phi_1 +\eta '_2
 \Phi_1 \Phi_2 + \eta_3^{\prime} \Phi_1 \Phi_2^*)+...
 \eqn\nonrenormsofttril$$
We have omitted $SU(2)_H$ indices on $Q$, $\bar u$, $\bar d$ but terms
with all possible contractions should be understood.  Here $\gamma$,
$\gamma^{\prime}$, $\eta$ and $\eta^{\prime}$ are dimensionless
numbers.  The couplings $\eta_3$ and $\eta_3^{\prime}$ may seem
surprising, since they are not among the usual allowed soft breaking terms.
These terms, however, are {\it supersymmetric} terms, arising because
the superpotential of the effective theory will in general contain terms
like $m_{susy} \Phi_i \Phi_j$.
By 't Hooft's naturalness criterion,\refmark{\cargese}
many of the couplings in eqs. \softterms-\nonrenormsofttril\
should not be much less than one; the theory does not become
any more symmetric if these quantities vanish.  Some, however,
can naturally be small; later, we will consider discrete symmetries
which might suppress certain dangerous ones.

The resulting down squark mass matrices are of three types;
$\tilde{m}^2_{LL}$, $\tilde{m}^2_{LR}$, and $\tilde{m}^2_{RR}$, where
L and R refer to left-handed and right-handed squarks, respectively.
For example, for $\tilde{m}^2_{LL}$ one obtains
$$\tilde{m}^2_{LL}=diag(\tilde m_1^2, \tilde m_1^2, \tilde m_2^2)
+\delta \tilde{m}^2. \eqn\diagsquark$$
The first and second terms originate in $V_{soft}$ and
$\delta V_{soft}^2$,
respectively.  In the interaction basis, the (13), (23),
(31) and (32) entries of $\delta \tilde m^2$ are proportional to $\phi \over
M_p$
and the remaining entries are proportional to ${\phi^2 \over M_p^2}$.
In the quark mass eigenstate basis, the same is true (see
eqs. \downeigenstates).  For example, $\tilde m^2_{ds}$ is given by
$$\tilde m_{ds}^2=(\tilde m_2^2-\tilde m_1^2) x_{13}^* x_{23}
+ \delta \tilde m_{12}^2 +
(\delta\tilde m_{11}^2-\delta \tilde m_{22}^2) x_{21}
+ \delta \tilde m_{13}^2 x_{23} + \delta \tilde m_{32}^2 x_{13}^*+...
\eqn\llinsertions$$
% \xij, the off-diagonal entries are given by
%$$\tilde{m}^2_{ds}=(m_2^2 -m_1^2)x_{13}^* x_{23}
%+ \delta \tilde{m}^{ 2}_{12}
%+(\delta \tilde{m}^2_{11}-\delta \tilde{m}^2_{22})x_{21}+...\sim
%m_{susy}^2
%{\cal O}\left (\gamma ' {\phi^2 \over M_p^2}\right)$$
%$$\tilde{m}^2_{db}= (m_2^2 -m_1^2)x_{13}^* + \delta \tilde{m}^{ 2}_{13}
%%+...\sim~ m_{susy}^2 {\cal O}\left (\gamma {\phi\over M_p}\right)
%\eqn\LLinsertions $$
%$$\tilde{m}^2_{sb}=(m_2^2 -m_1^2) x_{23}^* +\delta \tilde{m}^{ 2}_{23}
%+...\sim ~ m_{susy}^2 {\cal O}
%\left (\gamma {\phi\over M_p}\right).$$
Similar statements hold for
the $\tilde{m}^2_{RR}$  and for the up-sector.  The
matrix elements satisfy the promising hierarchy
$ \tilde{m}^2_{ds} <<  \tilde{m}^2_{db},  \tilde{m}^2_{sb}. $

What about proportionality of the quark and squark mass matrices?
The $\tilde m^2_{LR}$ matrix corresponding to a fermion matrix,
$m$, satisfying the hierarchy in eq. \downmatrix, is of the form
$$\tilde m^2_{LR} \approx
\left ( \matrix{A_{11} m_{11} & A_{12} m_{22} & A_{13} m_{23} \cr
A_{21} m_{22} & A_{22} m_{22} & A_{23} m_{23} \cr
A_{31} m_{32} & A_{32} m_{32} & A_{33} m_{33}} \right ) .\eqn\proportion$$
For a general potential, the violations of proportionality for
the $\tilde{m}^2_{LR}$ matrix for the first two generations are of
order $\Phi^2/M_p^2$.  This is small enough for the $K-\bar K$
and $D-\bar D$ systems.

We now discuss implications for FCNC's.
Bounds on off-diagonal down squark masses \refmark{\masiero}
from the $K-\bar K$ and $B-\bar B$ mass differences are summarized
in Table 1.  Estimates of these
quantities in the $SU(2)_H$ model are collected in Table 2.
Similar estimates are obtained in the up-sector for $\tilde m^2_{LL}$
and $\tilde m^2_{RR}$, but  $\tilde m^2_{LR}$ entries will be related to the
up quark mass matrix and will be somewhat larger.  This will turn
out to be of significance for $D-\bar D$ mixing.
\REF\giudice{R. Barbieri and G.F. Giudice, {\sl Nucl. Phys.} {\bf B306}
(1988) 77; G.G. Ross and R.G. Roberts, {\sl Nucl. Phys.} {\bf B377} (1992)
571; B. de Carlos and J.A. Casas, CERN-TH. 6835/93.}
We see that for $\sim$ 300 GeV squarks and gluinos, some of the
dimensionless couplings in $\delta V^2_{soft}$ will have to be
$\sim {1\over 10} - {1\over 3}$ to obtain satisfactory $\Delta m_K$ and
$\Delta m_B$.  If the squarks of the first two generations have masses of
order 1 TeV then all dimensionless couplings in the scalar potential
can be ${\cal O}(1)$.  Provided the third generation squarks are
 comparatively light\refmark{\giudice}
(perfectly possible in this sort of model), this
does not imply any fine tuning.
Alternatively, as will be described below, discrete symmetries
can further suppress the most dangerous couplings.

We have not included limits from ${\rm Br}(b \to s \gamma)$ in Table 1.
{}From gluino graphs with LR squark mass
insertions \refmark{\masiero} the new CLEO bound of $5.4 \times 10^{-4}$
implies \foot{Graphs with LL and RR insertions give small contributions,
except for very light $O(100~GeV)$ squarks,
which are disfavored by $K-\bar K$ bounds.}
$$\left (\tilde{m}_{b \bar s}^2 \over \tilde{m}^2 \right)^2
{1\over \tilde {m }^2} \lsim 8 \times 10^{-10}, \eqn\bsgamma$$
and the same for $\tilde{m}_{s \bar b}^2$.
The $SU(2)_H$ model gives
$$\tilde{m}^2_{b \bar s}= A^d_{32} m^d_{32}
+\bar x^d_{23}A^d_{33}m_{33}^d+....$$
where the $A^d$'s are of order $m_{susy}$ as given in eq. \proportion.
For $m_{32} \sim m_s$, as in eq. \downmatrix,
this is too small to give interesting
contributions to $b \to s \gamma$.  However, if $m_{32}^d$
is as large as a few GeV but less than $m_b$, which is allowed from the point
of view of fermion masses and mixing angles, then, from eq. \bsgamma,
we see that
gluino graphs can contribute to ${\rm Br}(b\to s\gamma)$ at the level of
$5.4 \times 10^{-4}$ for squarks as heavy as 300 GeV; as noted
earlier, however, this may lead to difficulties for
$K -\bar K$.
\REF\bigi{L.I. Bigi and F. Gabbiani, {\sl Nucl. Phys.} {\bf B352} (1991) 309.}
\REF\hqet{H. Georgi, Harvard preprint HUTP-92/A049; T. Ohl, G. Ricciardi and
E.H. Simmons, Harvard preprint HUTP-92/A053.}

Since the bounds from $\Delta m_B$ are just barely satisfied, the model
could have very rich implications for $CP$-violation in the $B$
system.\refmark{\bigi}
As remarked above, the $\tilde m^2_{LR}$
entries have interesting consequences for $D-\bar D$ mixing.
Given the current experimental bound of
${\Delta m_D \over m_D }< 6.97 \times 10^{-14}$ one obtains the
following constraints,  taking $F_D=200~MeV$ and $B_D=1$:
$$\left({\tilde{m}^2_{u\bar c}
\over\tilde{m}^2} \right)^2 {1\over \tilde{m}^2}
\lsim 10^{-9} \;{\rm GeV}^{-2}\eqn\ddbar$$
for $\tilde m_g \sim \tilde m$; the bound is $7 \times 10^{-10}$
for $\tilde m_g \sim 0.1\;\tilde m$.
By way of comparison, in the $SU(2)_H$ model we expect for the above quantities
$2\times 10^{-10}\eta '^2$ for 300 GeV squark masses, and
$2 \times 10^{-12}\eta'^2$ for $\sim 1$ TeV squark masses.  Recent
HQET calculations \refmark{\hqet} for the standard model lead
to ${\Delta m_D \over m_D }\sim 10^{-17} - 10^{-16} $.  So it is clear that
one can readily obtain $\Delta m_D$ one to two orders of magnitude larger
in the $SU(2)_H$ model!

What about constraints from $CP$-violation?  The bounds from
$\epsilon _K$ on the imaginary parts of the various quantities
constrained by $\Delta m_K$ are about two orders of magnitude
stronger than indicated in Table 1.  Constraints on non-degeneracy
will depend on the size of the phases entering these quantities.
For example, if the phases are of order unity,
as would be expected in models with explicit $CP$-violation,
then for $\sim 1$ TeV first and second generation
squarks, some of the $\gamma '$
couplings in $\delta V_{soft}$ would have to be of order ${1\over 10}$.
Couplings of this size or smaller might arise
as a consequence of discrete symmetries, as discussed below.
Alternatively, these squarks could be even heavier.
This will not lead to fine-tuning of Higgs parameters, since, as
already mentioned, their Yukawa couplings
are small.  Note that with this choice of parameters, large $CP$-violating
supersymmetric contributions in the $B$-system are possible, since
the relevant phases can be of order unity.

\REF\nedm{J. Ellis, S. Ferrara and D.V. Nanopoulos, {\sl Phys. Lett.} {114B}
(1982) 231; J. Polchinski and M. Wise, {\sl Phys. Lett.} {\bf 125B} (1983)
393; W. Buchmuller and D. Wyler, {\sl Phys. Lett.} {\bf 121B} (1983) 321.}
The bound on the neutron electric dipole moment, $d_n$,
most strongly constrains $Im (m_{\tilde g}A_{11}m^q_{11})$,
and $Im ({\tilde{m}^2_{u \bar t} }{\tilde{m}^2_{ t \bar u}}A^u_{33}m^u_{33})$.
The former constraint has been widely considered\refmark{\nedm} in the MSSM.
The situation here is much the same: for $\sim300$ GeV squark,
gluino and $A_{ij}$ trilinear scalar masses,
one requires $Arg [m_{\tilde g}A_{11}m^q_{11}] \sim 10^{-2}$,
while for $\sim 1$ TeV squarks, phases of order unity are permissible.
The same turns out to be true for $Arg[\tilde{m}^2_{u \bar t}
\tilde{m}^2_{t\bar u}A^u_{33}m^u_{33}]$, given that the $B-\bar B$
constraints on $\tilde m^2_{LR}$
are satisfied in our model.

\REF\pomarol{A. Pomarol, {\sl Phys.Rev.} {\bf D47} (1992) 273 ;
R. Garisto and G. Kane, TRIUMF preprint TRI-PP-93-1, (1993); L. Hall and
S. Weinberg, Texas preprint UTTG-22-92.}
\REF\strominger{A. Strominger and E. Witten, {\sl Comm. Math. Phys.}
{\bf 101} (1985) 341.}
In scenarios of spontaneous $CP$-violation, the relevant phases might
naturally be of order $10^{-2}$, in which case the $\epsilon _K$,
and $d_n$ constraints are more comfortably accomodated\refmark{\pomarol}.
One interesting possibility is that it is the
$\Phi$ field vev's which spontaneously break $CP$.  This
is consistent, for example, with the idea described above
that these fields could be moduli of a string
compactification.\refmark{\strominger}  In such
a scheme the phase of the gluino, in particular, will be at most of
order $\Phi^2 /M_p^2$.    However, it appears difficult to suppress
$Arg \langle\Phi \rangle$, so that the other phases of relevance to
$d_n$ and $\epsilon _K$ are likely to be large.
Perhaps if the scale of $CP$-violation is somewhat smaller than the
scale of horizontal symmetry breaking one can naturally obtain smaller phases.

An alternative strategy for accomodating $CP$-violation bounds
is to increase the amount of squark degeneracy by adding additional
abelian discrete or continuous horizontal symmetries.
One notices that all terms in eqs. \nonrensoftbil\ and \nonrenormsofttril\
which contribute to off-diagonal entries (here we are refering to the
interaction basis) in $\tilde{m}^2_{LL}$, $\tilde{m}^2_{RR}$ and
$\tilde{m}^2_{LR}$ can, in principle, be eliminated by additional
symmetries. It is not hard to construct models with such symmetries
and realistic quark mass matrices. The smallness of off-diagonal
squark mass matrix entries is limited by the $x_{ij},~\bar x_{ij}$,
or KM angles. This can lead to further suppression of order $\theta_c^2$
for $K-\bar K$ and of order $\left({V_{cb}\over \gamma}\right)^2$
for $B-\bar B$. Moreover, in many cases, the lowest dimension operators
are $CP$-conserving, providing adequate suppression for
Im $K^o-\bar K^o$.

One can also carry out the above program making use of other
symmetry groups, such as $SU(3)$ or non-abelian discrete
groups. The SU(2) models have the virtue of simplicity, which is in
large part due to the gross features of the quark mass
spectrum:  large mass splitting
and small mixing angles between the third family and the first two.

Let us turn now to the possibility of breaking at lower scales.
We will not attempt here to construct explicit models, but confine
ourselves to some general remarks.  First, there are a number
of approaches one might adopt.
We have already remarked that if the Higgs carry $SU(2)_H$ quantum
numbers, there are likely to be problems with flavor
changing neutral currents from Higgs exchange.  Still, such models
are clearly worthy of exploration.

An alternative possibility, following ref. \nir, is to suppose that at
some new scale, not far from the flavor
symmetry breaking scale, there are
some $SU(2)_L$-singlet vector-like quarks, some of which are in doublets of
$SU(2)_H$.  There
are also $SU(2)_H$-breaking doublets $\phi_i$, like in the large-scale
model, which couple the light and heavy quarks.  Without a terribly
complicated structure at this scale, integrating out the heavy
fields produces couplings of light quarks analogous to those in eq.
\dimfiveandsix\ with $M_p$ replaced by the heavy quark
scale.

We can summarize all of this by saying that it is
easy to construct models in which horizontal symmetries
adequately suppress flavor changing neutral currents.
This view suggests patterns of masses which may
differ from assumptions which are conventional
in model building.  For example, at very high energies,
the third generation (left and right)
squarks are not likely to be degenerate
with those of the first two.  Moreover, in the simplest
models, the squarks of the
first two generations should have masses of order a TeV,
while the top squark (to avoid naturalness problems)
should be comparatively light.
The simplest models, which only make use of an $SU(2)$ horizontal
symmetry, offer no understanding
of the quark mass matrix.  Such an understanding
may require more intricate symmetry patterns, which,
as we have illustrated, may lead to even tighter degeneracy.
Perhaps, after all, supersymmetry may yield insights
into the problems of flavor.

\centerline{\bf Acknowledgements}
We thank N. Seiberg for discussions of many of the issues considered here,
and N. Seiberg and Y. Nir for helpful comments on an early version of the
manuscript.
The work of M.D. and R.L was supported in part by the U.S. Department of
Energy under contract \#DE-FG03-92ER40689, and that of A.K. by
contract \#DE-AC03-76SF00515.
\refout
\endpage
%Tables follow. Uses tables.tex.

\vbox{\tenrm
  {\narrower\noindent%
{\bf Table 1a}.\ \ %
Bounds on various components of the squark mass matrix from
$K^o -\bar K^o$ mixing. Here, $m_{\tilde g}$ is the gluino
mass and $\tilde m$ is a typical first-second generation squark
mass. All quantities in GeV$^{-2}$. We have taken $F_K=170$ MeV
and $B_K=1$.
\smallskip
}%
\medskip
\begintable
${m_{\tilde g} \over \tilde{m}}$ | ${\tilde {m}_{ds}^4 \over
\tilde{m}^6}$ | ${\tilde{m}_{ds}^2 \tilde{m}_{\bar d \bar s}^2 \over
\tilde{m}^6}$ | ${\tilde{m}_{d \bar s}^4 \over \tilde {m}^6}$ \cr
$1$ | $2 \times 10^{-9}$ | $3 \times 10^{-11}$ |$ 10^{-10}$ \cr
$.1$ | $4 \times 10^{-10}$ | $10^{-10}$ | $2 \times 10^{-11} $ \endtable
}
\vskip1cm
\vbox{\tenrm
  {\narrower\noindent%
{\bf Table 1b}.\ \ %
Bounds on various components of the squark mass matrix from
$B^o -\bar B^o$ mixing. We have taken $F_B=230$ MeV and $B_B=1$.
\smallskip
}%
\begintable
${m_{\tilde g} \over \tilde{m}}$ | ${\tilde {m}_{db}^4 \over
\tilde{m}^6}$ | ${\tilde{m}_{db}^2 \tilde{m}_{\bar d \bar b}^2 \over
\tilde{m}^6}$ | ${\tilde{m}_{d \bar b}^4 \over \tilde {m}^6}$ \cr
$1$ | $ 10^{-8}$ | $ 10^{-9}$ |$2 \times 10^{-9}$ \cr
$.1$ | $2 \times 10^{-9}$ | $2 \times 10^{-9}$ | $7 \times 10^{-10} $
 \endtable
}
\endpage
\vskip2.5cm
\vbox{\tenrm
  {\narrower\noindent%
{\bf Table 2a}.\ \ %
Predictions of the $SU(2)_H$ model for $K^o -\bar K^o$ mixing
for representative values of squark masses. The dimensionless
parameters $\gamma$, $\eta$, \etc, in this and the following
table are as defined in the text. All quantities in GeV$^{-2}$.
\smallskip
}%
\medskip
\begintable
$ \tilde{m}$ | ${\tilde {m}_{ds}^4 \over
\tilde{m}^6}$ | ${\tilde{m}_{ds}^2 \tilde{m}_{\bar d \bar s}^2 \over
\tilde{m}^6}$ | ${\tilde{m}_{d \bar s}^4 \over \tilde {m}^6}$ \cr
$ 300~GeV$ | $ 10^{-9} \gamma '^2 $ | $ 10^{-9} \gamma '^2 $ |$ 3
\times 10^{-12} \eta '^2$ \cr
$ 1~TeV$ | $ 10^{-10} \gamma '^2 $ | $ 10^{-10} \gamma '^2 $ |$ 3
\times 10^{-14} \eta'^2 $ \endtable
}

\vskip1cm
\vbox{\tenrm
  {\narrower\noindent%
{\bf Table 2b}.\ \ %
Predictions of the $SU(2)_H$ model for $B^o -\bar B^o$ mixing
for representative values of squark masses.
\smallskip
}%
\smallskip
\begintable
$ \tilde{m}$ | ${\tilde {m}_{db}^4 \over
\tilde{m}^6}$ | ${\tilde{m}_{db}^2 \tilde{m}_{\bar d \bar b}^2 \over
\tilde{m}^6}$ | ${\tilde{m}_{d \bar b}^4 \over \tilde {m}^6}$ \cr
$ 300~GeV$ | $ 10^{-7} \gamma ^2 $ | $ 10^{-7} \gamma ^2 $ |$ 3
\times 10^{-12} \eta^2$ \cr
$ 1~TeV$ | $ 10^{-8} \gamma ^2 $ | $ 10^{-8} \gamma ^2 $ |$ 3
\times 10^{-14} \eta^2 $ \endtable
}
\bye